\documentclass[aps,prl,reprint,amsmath,amssymb,superscriptaddress]{revtex4-1}
\usepackage{epsfig}
\usepackage{graphicx, subfigure}
\usepackage{color}
\usepackage{amsfonts}
\usepackage{amsmath}
\usepackage{amssymb}
\usepackage{amsthm}
\usepackage{braket}
\usepackage{multirow}
\usepackage{array}
\usepackage{tabularx}

\usepackage{lipsum}

\newcommand{\kb}{\mathbf{k}}

\begin{document}

\title{Spin-polarized current in non-collinear antiferromagnets} 

\author{Jakub \v{Z}elezn{\'y}}
\affiliation{Max Planck Institute for Chemical Physics of Solids, 01187 Dresden, Germany}

\author{Yang Zhang}
\affiliation{Max Planck Institute for Chemical Physics of Solids, 01187 Dresden, Germany}
\affiliation{Leibniz Institute for Solid State and Materials Research, 01069 Dresden, Germany}

\author{Claudia Felser}
\affiliation{Max Planck Institute for Chemical Physics of Solids, 01187 Dresden, Germany}

\author{Binghai Yan}
\affiliation{Max Planck Institute for Chemical Physics of Solids, 01187 Dresden, Germany}
\affiliation{Department of Condensed Matter Physics, Weizmann Institute of Science, 7610001 Rehovot, Israel}

\begin{abstract}
Noncollinear antiferromagnets, such as Mn$_3$Sn and Mn$_3$Ir, were recently shown to be analogous to ferromagnets in that they have a large anomalous Hall effect. Here we show that these materials are similar to ferromagnets in another aspect: the charge current in these materials is spin-polarized. In addition, we show that the same mechanism that leads to the spin-polarized current also leads to a transverse spin current, which has a distinct symmetry and origin from the conventional spin Hall effect. We illustrate the existence of the spin-polarized current and the transverse spin current by performing \emph{ab initio} microscopic calculations and by analyzing the symmetry. We discuss possible applications of these novel spin currents, such as an antiferromagnetic metallic or tunneling junction.
\end{abstract}

\maketitle

\paragraph{Introduction.}
Spintronics is a field that studies phenomena in which both spin and charge degree of electron play an important role. Many of the key spintronics effects are based upon the existence of spin currents. Two main types of spin currents are utilized: the spin-polarized currents in ferromagnets (FMs) and the spin currents due to the spin Hall effect (SHE) which are transveral to the charge current and appear even in non-magnetic materials. The most important effects that originate from the spin-polarized currents in FMs are the giant and the tunneling magnetoresistance effects (GMR and TMR) \cite{JULLIERE1975225,Miyazaki1995,Moodera1995} and the spin-transfer torque (STT) \cite{Slonczewski1989,Ralph2008}. These effects are utilized for magnetic sensing and in the magnetic random access memories (MRAMs) \cite{Khvalkovskiy2013}. This memory is non-volatite and has speed and density comparable to the widely used dynamic random access memory. The SHE is pivotal for spintronics since it allows transforming charge current into a spin current. It is responsible (though other effects can contribute) for the spin-orbit torque (SOT) \cite{Miron2011,Liu2012} in multilayer heterostructures, which can be used for efficient and fast switching of FM layers. The SOT is now also being explored for use in MRAMs \cite{Prenat2016,Oboril2015}.

While spintronics has traditionally utilized FM and non-magnetic materials, in the past few years also antiferromagnetic (AFM) materials have attracted a considerable interest. AFMs offer some unique advantages compared to FMs, but are much less explored (see reviews \cite{Macdonald2011,Jungwirth2015,Baltz2016}). AFMs have a very fast dynamics, which allows for switching on ps timescale \cite{Gomonay2016,Roy2016,Kimel2004}. Furthermore, there exists a wide range of AFM materials, including many insulators and semiconductors, multiferroics \cite{Eerenstein2006} and superconductors \cite{Lu2015}. Utilizing (and also studying) AFMs is difficult, largely because the magnetic order in AFMs is hard to detect and to manipulate.

AFM spintronics has so far focused mostly on collinear AFMs in which the electrical current is not spin-polarized. This limits the spintronics effects that can be observed in such AFMs. Here we show that this limitation only relates to the simple collinear AFMs. We demonstrate by means of symmetry arguments and ab-inito calculations that in non-collinear AFMs novel type of spin currents occur. These spin currents have a longitudinal component (i.e., flowing along the same direction as the electrical current) or in other words the electrical current is spin-polarized. Unlike in FMs, these spin currents also have a large transverse component. Such a spin current resembles the SHE in that it is a spin current transverse to the charge current, however, it is fundamentally distinct from the SHE. A key distinction is that the spin currents we discuss here are odd under time-reversal, whereas the SHE is even. This is analogous to the distinction between normal current and the anomalous Hall effect (AHE).

Successful experimental demonstrations of electrical detection and manipulation of AFMs has utilized relativistic effects which do not rely on the spin-polarized current \cite{Park2011,Fina2014,Wang2014,Kriegner2016,Wadley2016,Zelezny14,Wadley2016}. These methods could be used to develop AFM spintronic devices, but they have some disadvantages compared to the methods used in FMs. Our work shows that in non-collinear AFMs spintronics could instead be developed along a similar route as FM spintronics. As an example we propose that a magnetoresistance and STT will be present in an AFM junction.

The transverse spin currents are also important for spintronics as they allow for similar functionality as the SHE, but have a different origin and symmetry. This could, for example, be useful for the SOT since the high symmetry of SHE in commonly used metals is limiting \cite{MacNeill2016}. Additionally, the odd spin currents are directly relevant for experiments which demonstrated a large SOT in non-collinear AFM/FM heterostructures \cite{Tshitoyan2015,Reichlova2015,Zhang2016,Oh2016,Wu2016}.

\begin{figure}[h]
  \centering
  \includegraphics[width=8.6cm]{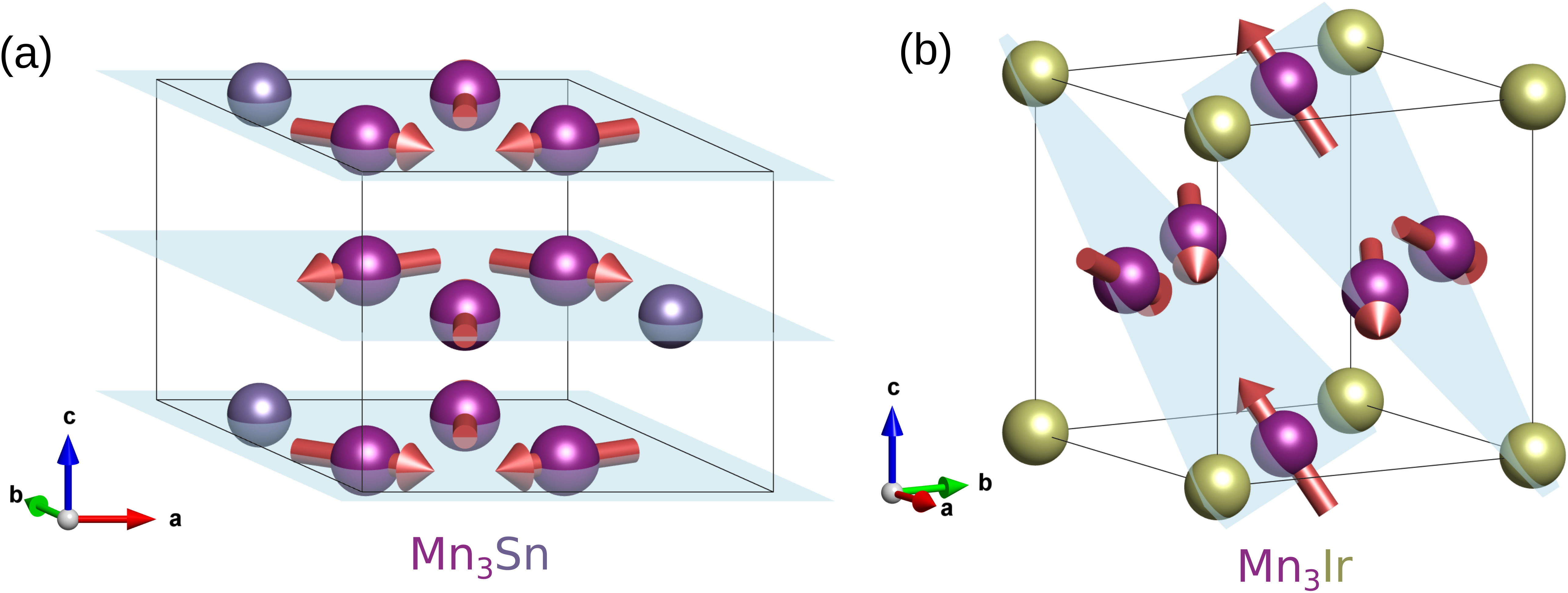}
  \caption{(a) Crystal and magnetic structure of Mn$_3$Sn (as well as Mn$_3$Ga and Mn$_3$Ge) and (b)  Mn$_3$Ir (as well as Mn$_3$Rh, Mn$_3$Pt).}
  \label{fig:structures}
\end{figure}

We illustrate the existence of the novel spin currents on non-collinear AFMs Mn$_3$Sn and Mn$_3$Ir, which have triangular magnetic configurations shown in Figs \ref{fig:structures}(a),(b). These AFMs have recently attracted attention because they were shown to have a large anomalous Hall effect (AHE) \cite{Chen2014,Kubler2014,Nakatsuji2015,Nayak2016} as well as a magneto-optical Kerr effect \cite{Feng2015}, even though they have only a very small net magnetization (which is not the origin of these effects). The conventional SHE in the non-collinear AFMs was already theoretically studied in depth in Ref. \cite{YangZhang2017}, thus we focus here only on the odd spin currents. 

\paragraph{Symmetry analysis and calculations.}

The response of metals to electric fields can be described well by linear response theory. Here we use the so-called constant $\Gamma$ approximation, i.e., we assume that the only effect of disorder is a constant band broadening, which modifies the Green's functions of the perfectly periodic system in the following way:
$ G^{R}(\varepsilon) = 1/(\varepsilon-\hat{H} + i 0+)  \rightarrow 1/(\varepsilon-\hat{H} + i\Gamma),$
where $\hat{H}$ is the Hamiltonian, $\varepsilon$ is energy, $G^R$ is the retarded Green's function (the advanced Green's function is modified analogously) and $\Gamma$ is a constant that determines the broadening magnitude. Every linear response formula can be decomposed into two contributions which transform in the opposite way under time-reversal. Within the constant $\Gamma$ approximation the two contributions to the linear response of an observable $\hat{A}$ to an electric field are given by $\delta{A} = \chi^I E+\chi^{II}E$, where \cite{freimuth2014}

\begin{align}
 \chi^{I} &= -\frac{e\hbar}{\pi}\sum_{\kb, n,m} \frac{\Gamma^2 \text{Re}\left(\Bra{n\kb} \hat{A} \Ket{m\kb} \Bra{m\kb} \hat{\mathbf{v}} \cdot \hat{\mathbf{E}} \Ket{n\kb}\right)}{[(E_F-\varepsilon_{n\kb})^2+\Gamma^2][(E_F-\varepsilon_{m\kb})^2+\Gamma^2]}, \label{eq:Boltzmann}\\
  \chi^{II}  &= -2\hbar e 
  \sum_{\substack{\kb,n \neq m}} ^{\substack{ \\ n\ \text{occ.}\\ m\ \text{unocc.}}}
  \frac{\text{Im}\left(\Bra{n\kb}\hat{A}\Ket{m\kb}\Bra{m\kb}\hat{\mathbf{v}}\cdot \hat{\mathbf{E}}\Ket{n\kb}\right)}{(\varepsilon_{n\kb} - \varepsilon_{m\kb})^2},\label{eq:intrinsic}
\end{align}
Here $e$ is the (positive) elementary charge, $\kb$ is the Bloch wave vector, $n,m$ are the band indices, $\varepsilon_{n\kb}$ is the band energy, $E_F$ is the Fermi energy, $\hat{\mathbf{v}}$ is the velocity operator, $\hat{\mathbf{E}}$ is the direction of the electric field and $E$ its magnitude. In Eq. \eqref{eq:intrinsic} the sum is restricted to $m,n$ such that $n$ is occupied and $m$ is unoccupied. The sums over $\kb$ run over all $\kb$ points in the first Brillouin zone. We give here the contribution $\chi^{II}$  only in the limit $\Gamma \rightarrow 0$, as this expression is usually considered in this limit. This contribution is known as the intrinsic contribution because it is determined only by the electronic structure of the perfect crystal. In the limit $\Gamma \rightarrow 0$, Eq. \eqref{eq:Boltzmann} becomes the well known Boltzmann formula with constant relaxation time (with the relaxation time given by $\hbar/2\Gamma$). This contribution is diverging as $1/\Gamma$ when $\Gamma \rightarrow 0$.  While these formulas are simple they often provide at least qualitatively and sometimes even quantitatively correct description. We use them to illustrate the symmetry of linear response and to confirm the existence of the novel spin currents. When $\hat{A}$ is equal to current density operator: $\hat{\mathbf{j}} = -e\hat{\mathbf{v}}/V$, Eqs. \eqref{eq:Boltzmann}, \eqref{eq:intrinsic} describe conductivity. When $\hat{A}$ is set to the spin-current operator, $\hat{j}^s_{i,j} = \frac{1}{2}\{\hat{s}_i,\hat{v}_j\}$, these equations instead describe spin-conductivity.

Eqs. \eqref{eq:Boltzmann} and \eqref{eq:intrinsic} transform differently under time-reversal because time-reversal is an anti-unitary operator, which transforms the matrix elements as: $\Bra{n\kb}\hat{A}\Ket{n\kb} \rightarrow \Bra{n\kb}T\hat{A}T\Ket{n\kb}^*$ \cite{Zelezny2017}. Because of the complex conjugation the term \eqref{eq:intrinsic} will contain additional minus under a time-reversal transformation compared to the term \eqref{eq:Boltzmann}. Thus for conductivity the term \eqref{eq:Boltzmann} is even under time-reversal, while the term \eqref{eq:intrinsic} is odd. Note that equivalently these terms are also even resp. odd under the reversal of all magnetic moments. The even part describes the ordinary conductivity, while the odd part describes the AHE. Since AHE is odd under time-reversal it can be nonzero only in a magnetic system (assuming non-interacting electrons). Traditionally, it has been considered for FMs only, but recently it was shown that Eq. \eqref{eq:intrinsic} is also nonzero and relatively large in non-collinear AFMs \cite{Kubler2014,Chen2014}. Collinear AFMs are typically symmetrical under simultaneous time-reversal and lattice translation or under simultaneous time-reveral and inversion and these symmetries prohibit the existence of AHE. In non-collinear AFMs, both of these symmetries are usually broken and thus the non-collinear AFMs can in general have an AHE. 

For spin-conductivity the transformation under time-reversal is precisely opposite because the spin current operator contains an additional spin operator which is odd under time-reversal. Thus for spin-conductivity, Eq. \eqref{eq:Boltzmann} is odd under time-reversal, while Eq. \eqref{eq:intrinsic} is even. The spin-currents that are even under time-reversal are known as the SHE. The odd spin-currents were previously considered only in FMs, however, as we will show in this manuscript they also exist in non-collinear AFMs (while in collinear AFMs they will typically be prohibited by the same symmetries as AHE), in complete analogy to the AHE. Since the intrinsic contribution to the spin currents in the triangular AFMs was recently explored in detail in \cite{YangZhang2017}, we focus here only on the spin-currents that are odd under time-reversal described by Eq. \eqref{eq:Boltzmann}. To evaluate this equation, the ground state eigenvalues and eigenfunctions are needed, which we obtain from a non-collinear density functional theory calculation. We use the VASP code with the PBE-GGA exchange-correlation potential. To make the calculation faster we utilize the Wannier interpolation \cite{freimuth2014,YangZhang2017}; see the Supplemental Material \cite{SMprl} for a detailed description of the method.

Within linear response we can describe the spin current using a spin-conductivity tensor $\sigma^i_{jk}$, such that $\sum_k \sigma^i_{jk} E_k$ is the spin current with spin-polarization along $i$ and flowing in the direction $j$. By considering all the symmetry operations and how they transform the spin-conductivity tensor \cite{Zelezny2017,Seeman2015,symcode} we find that the odd spin currents are indeed allowed by symmetry in the Mn$_3$X compounds. Note that this symmetry analysis is not related to the constant $\Gamma$ approximation, but applies generally for any linear response. In Table \ref{tab:symmetry} we give the general form of the odd spin-conductivity tensors for Mn$_3$Sn and Mn$_3$Ir. These symmetry tensors presume the existence of SOC. We find that the spin currents in the triangular AFMs appear even without the SOC (note that this is also true for the SHE in Mn$_3$Sn \cite{Zhang2017arxiv}). In absence of the SOC, the symmetry is higher because spin is then not coupled to the lattice directly. The symmetry restricted shape of the odd spin-conductivity tensors in absence of SOC is also given in Table \ref{tab:symmetry}. These tensors can be derived by considering combination of symmetries of the nonmagnetic lattice with pure spin rotations \cite{Litvin1974,Brinkman1966} and are in good agreement with our calculations. See the Supplemental Material \cite{SMprl} for further details. As shown in  Table \ref{tab:symmetry}, the symmetry in absence of the SOC is much higher than with the SOC. Both in Mn$_3$Sn and in Mn$_3$Ir the $\sigma^i_{jk}$ tensors have only one independent component without the SOC. 

\begin{table}
\centering
\begin{ruledtabular}
 \begin{tabular}{cccc}
  && no SOC & SOC \\
  \hline
      \noalign{\vskip 2pt}   
    & $\sigma^x$& $\left(\begin{matrix} 
       0 & \sigma^x_{xy} & 0 \\
       \sigma^x_{xy} & 0 & 0 \\
       0 & 0 & 0
      \end{matrix}\right) $
 & $\left(\begin{matrix}0 & \sigma^x_{xy} & 0\\\sigma^x_{yx}& 0 & 0\\0 & 0 & 0\end{matrix}\right)$ \\
 Mn$_3$Sn &$\sigma^y$& $\left(\begin{matrix} 
       -\sigma^x_{xy}& 0 & 0 \\
       0 & \sigma^x_{xy} & 0 \\
       0 & 0 & 0
      \end{matrix}\right) $
   & $\left(\begin{matrix}\sigma^y_{xx} & 0 & 0\\0 & \sigma^y_{yy} & 0\\0 & 0 & \sigma^y_{zz} \end{matrix}\right)$ \\
    &$\sigma^z$& $\left(\begin{matrix} 
       0 & 0 & 0 \\
       0 & 0 & 0 \\
       0 & 0 & 0
      \end{matrix}\right) $
   & $\left(\begin{matrix}0 & 0 & 0\\0 & 0 & \sigma^z_{yz}\\0 &\sigma^z_{zy} & 0\end{matrix}\right)$ \\
      \noalign{\vskip 2pt}   
      \hline 
      \noalign{\vskip 2pt}   
   & $\sigma^x$ & $\left(\begin{matrix}\sigma^x_{xx} & 0 & 0\\0 & - \frac{\sigma^x_{xx}}{2} & 0\\0 & 0 & - \frac{\sigma^x_{xx}}{2}\end{matrix}\right)$
   & $\left(\begin{matrix}\sigma^x_{xx} & \sigma^x_{xy} & \sigma^x_{xy}\\\sigma^x_{yx} & \sigma^x_{yy} & \sigma^x_{yz}\\\sigma^x_{yx} & \sigma^x_{yz} & \sigma^x_{yy}\end{matrix}\right)$\\
   Mn$_3$Ir &$\sigma^y$ & $\left(\begin{matrix}- \frac{\sigma^x_{xx}}{2} & 0 & 0\\0 & \sigma^x_{xx} & 0\\0 & 0 & - \frac{\sigma^x_{xx}}{2}\end{matrix}\right)$
   & $\left(\begin{matrix}\sigma^x_{yy} & \sigma^x_{yx} & \sigma^x_{yz}\\\sigma^x_{xy} & \sigma^x_{xx} & \sigma^x_{xy}\\\sigma^x_{yz} & \sigma^x_{yx} & \sigma^x_{yy}\end{matrix}\right)$\\
   &$\sigma^z$ & $\left(\begin{matrix}- \frac{\sigma^x_{xx}}{2} & 0 & 0\\0 & - \frac{\sigma^x_{xx}}{2} & 0\\0 & 0 & \sigma^x_{xx} \end{matrix}\right)$
   & $\left(\begin{matrix}\sigma^x_{yy} & \sigma^x_{yz} & \sigma^x_{yx}\\\sigma^x_{yz} & \sigma^x_{yy} & \sigma^x_{yx}\\\sigma^x_{xy} & \sigma^x_{xy} & \sigma^x_{xx}\end{matrix}\right)$
 \end{tabular}
\end{ruledtabular}
\caption{Symmetry restricted form of the odd spin-conductivity tensors in Mn$_3$Sn and Mn$_3$Ir with and without SOC. For Mn$_3$Ir the tensors are given in the coordinate system shown in Fig. \ref{fig:structures}(b). For Mn$_3$Sn we use a cartesian coordinate system related to the coordinate system in Fig. \ref{fig:structures}(a) in the following way: $\mathbf{x}=\mathbf{a}$, $\mathbf{y}=(\mathbf{a}+2\mathbf{b})/\sqrt{3}$, $\mathbf{z} = \mathbf{c}$.}
\label{tab:symmetry}
\end{table}

\begin{figure}[h]
  \centering
  \includegraphics[width=8.6cm]{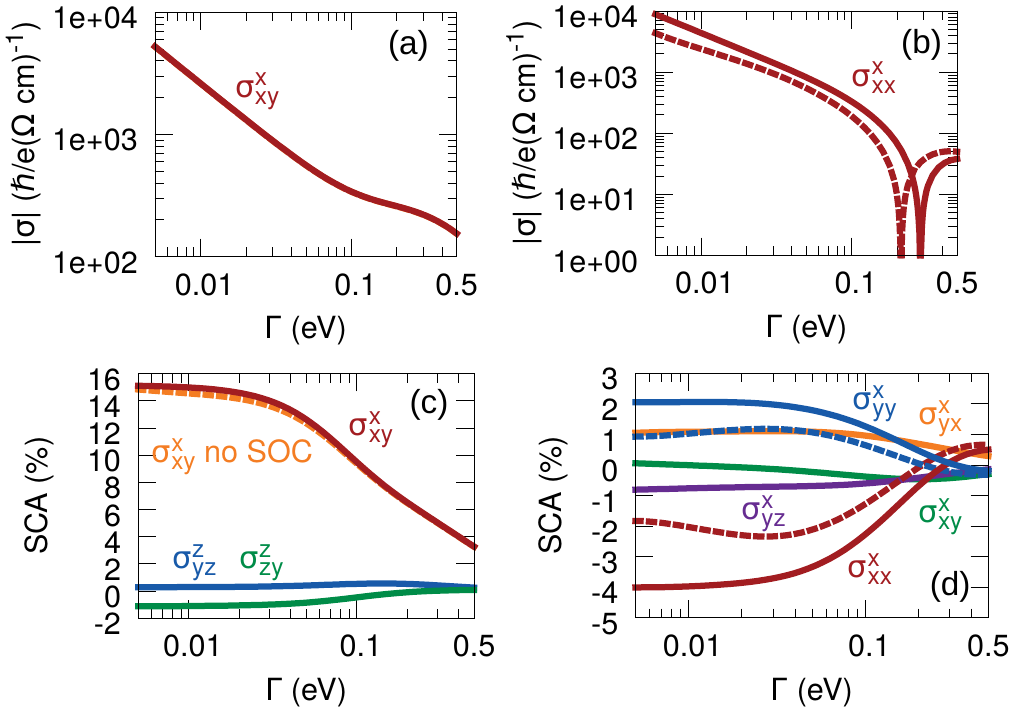}
  \caption{$\Gamma$ dependence of the odd spin currents in Mn$_3$Sn and Mn$_3$Ir.  (a),(b) The magnitude of the spin current for Mn$_3$Sn and Mn$_3$Ir respectively. Only the largest component is shown for clarity. The dashed line denotes a calculation without SOC. The dip in (b) corresponds to a sign change. (c), (d) The SCA for Mn$_3$Sn and Mn$_3$Ir respectively. }
  \label{fig:gamma_dep}
\end{figure}

In Fig. \ref{fig:gamma_dep}(a),(b) we plot the dependence of the magnitude of the odd spin currents in Mn$_3$Sn and Mn$_3$Ir on $\Gamma$. As expected, for small $\Gamma$ the odd spin currents are diverging as $1/\Gamma$. 
The magnitude of the SHE is often given in terms of the spin Hall angle, which is defined as $\frac{e}{\hbar}\frac{\sigma^i_{jk}}{\sigma_{kk}}$, where $\sigma_{kk}$ is the conductivity \cite{Hoffmann2013}. Such quantity can be defined for any spin current. To distinguish it from the conventional spin Hall angle we call it the spin current angle (SCA). The SCA defined in this way is dimensionless. In Figs. \ref{fig:gamma_dep}(c),(d) we plot the SCA as a function of $\Gamma$ for Mn$_3$Sn and Mn$_3$Ir. To evaluate the SCA we calculated the conductivity using Eq. \eqref{eq:Boltzmann}. Since both the conductivity and the spin conductivity scale as $1/\Gamma$ for small $\Gamma$, the SCA is independent of $\Gamma$ for small $\Gamma$. As can be seen in Fig. \ref{fig:gamma_dep} we find that  large spin currents are present even in absence of the SOC. In the Supplemental Material \cite{SMprl} we give the calculation of the odd spin currents also for other Mn$_3$X compounds with the same structures as Mn$_3$Sn or Mn$_3$Ir.

We can estimate  the value of $\Gamma$ by comparing the calculated conductivity with the experimental conductivity. For Mn$_3$Ir the experimental conductivity at 300 K is $2.5\times 10^4\  (\Omega\cdot \text{cm})^{-1}$ \cite{Yamaoka1974}. This corresponds to $\Gamma \approx 0.05\ \text{eV}$. For Mn$_3$Sn we find that even for very large values of $\Gamma$ (up to $0.5\ \text{eV}$), the calculated conductivity is smaller than the experimental conductivity (see the Supplemental Material \cite{SMprl}). This is probably because real crystals contain a significant amount of disorder, which cannot be captured by the constant $\Gamma$ approximation.  

For comparison we calculated the odd spin currents in bcc Fe using the same method. We find that within the constant $\Gamma$ approximation the longitudinal SCA in Fe is $\sim18\%$ and the transverse SCA is $\sim 1\%$. Note that such a calculation is only a rough estimate because the spin-dependent scattering is very important in FMs. 

Both Mn$_3$Sn and Mn$_3$Ir are not fully compensated, but have a small magnetic moment. Since this magnetic moment is very small, it cannot explain the odd spin currents discussed here. This is confirmed by calculation for Mn$_3$Ir in which the net magnetic moment is set to zero.
%
%

\begin{figure}[h]
  \centering
  \includegraphics[width=8.6cm]{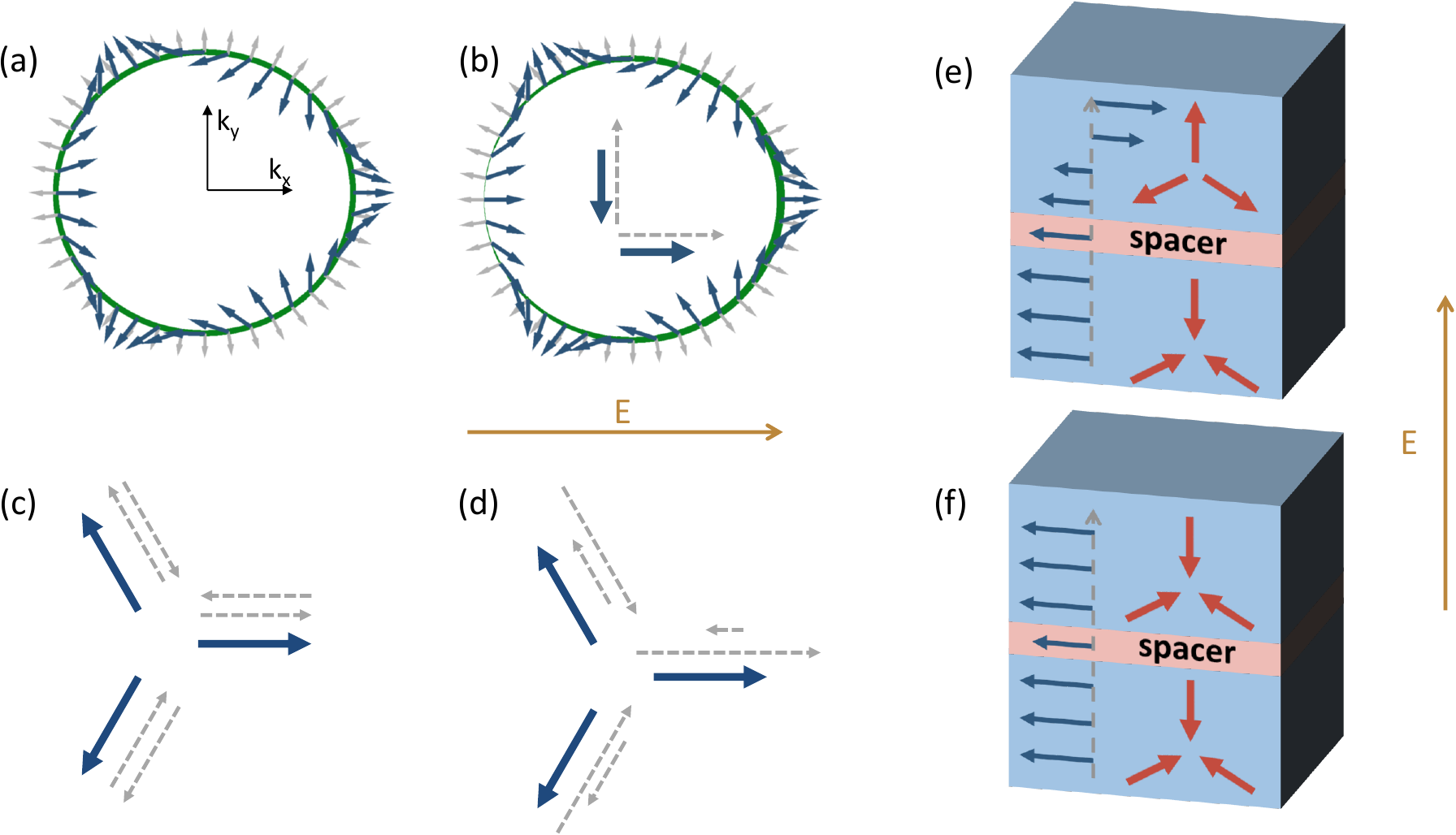}
  \caption{(a) Simplified Fermi level of a non-collinear AFM. Green line denotes the Fermi level, blue and gray arrows denote the mean values of spin and velocity respectively. (b) Electric field causes a redistribution of electrons at the Fermi level, signified by thicker or thinner green line. The arrows inside the circle show the corresponding spin currents. (c),(d) The main features of the Fermi level can be captured by considering only three types of electrons with velocities oriented parallel or antiparallel with their spin. (e),(f) Parallel and antiparallel states of the AFM junction. Gray dashed arrow denotes direction of the spin current flow, blue arrows denote the spin polarization of the spin current and red arrows denote the magnetic moments.  }
  \label{fig:TMR}
\end{figure}

\paragraph{Discussion.} 
The spin currents discussed here are similar to the spin-polarized currents in FMs, but they differ in some aspects. In FMs, in absence of SOC, spin is a good quantum number and the current can be decomposed into a spin-up and spin-down currents. This is the so-called two current model. Since the spin-up and spin-down electrons that carry the current have different properties (such as density, velocity or scattering rate), the spin-up and spin-down currents are different and the current is thus spin-polarized.

For non-collinear AFMs such description is not possible because in the presence of the non-collinear magnetic order, spin is not a good quantum number even without SOC. Therefore the electrons at the Fermi level can have spins oriented along various directions, as illustrated in Fig. \ref{fig:TMR}(a). Since there is no net magnetic moment, the integral of the spin of all electrons is zero. The integral of spin times velocity also vanishes and thus there is no spin current in equilibrium. Upon applying electric field, electrons at the Fermi level are redistributed (see Fig. \ref{fig:TMR}(b)). This results in a net current as well as a net spin current. The main features of the Fermi level depicted in Figs \ref{fig:TMR}(a),(b) can be captured by considering only three types of electrons, as illustrated in Figs. \ref{fig:TMR}(c),(d).  In is then easy to verify that the redistribution of electrons results in both longitudinal and transverse spin current (the resulting spin currents are shown in Fig. \ref{fig:TMR}(b)). In contrast, in FMs, in absence of SOC, the odd spin currents are only longitudinal. 

We first focus on the longitudinal spin currents. These spin currents are analogous to the spin-polarized currents in FMs and will thus have similar implications. When a spin-polarized current is injected into an AFM it generates a STT which can efficiently manipulate the AFM order \cite{Gomonay2010,Gomonay2012,Cheng2014c,Fujita2016arxiv}. Thus the STT will be present in a junction composed of two AFM layers separated by a thin metallic or insulating layer (see Figs. \ref{fig:TMR}(e),(f)). Such heterostructure is analogous to the FM spin valve or MTJ. With large enough current, the STT could be used to switch the junction between a parallel and an antiparallel configuration. Analogously to the case of SHE and inverse SHE, there must also exist an inverse effect to the spin-polarized current: a charge current generated by injection of a spin-polarized current. This current will flow in the opposite direction when the spin-polarization of the spin-polarized current is reversed. The parallel and antiparallel configurations will thus have a different conductivity or equivalently different resistance, similarly to the GMR or TMR effect. The AFM junction is thus in principle analogous in functionality to the FM spin valve or MTJ, however, predicting the magnitude of the magnetoresistance and the torque is beyond the scope of this work.

It has been predicted by many authors that magnetoresistance and a STT will occur even in spin valves or tunneling junctions composed of collinear AFMs in which current is not spin-polarized \cite{Nunez2006,Haney2007,Xu2008a,Haney2008a,Prakhya2014,Merodio2014,Saidaoui2014,Macdonald2011,Manchon2016,Saidaoui2016a,Stamenova2017}. These effects, however, rely on quantum coherence and perfect interfaces and were shown to be strongly suppressed by disorder \cite{Duine2007,Saidaoui2014,Manchon2016}. The effects we have described here, on the other hand, do not rely on perfect interfaces and are expected to be similarly robust as the analogous effects in FMs since they rely only on the existence of the spin-polarized current. We also remark that the longitudinal spin currents can occur in nonmagnetic materials as well if the crystalline symmetry is low enough \cite{Wimmer2015}. Such spin currents differ from the spin-polarized currents discussed here since they are even under time reversal and require SOC.

The transverse spin currents are similar to the spin currents due to the SHE, but differ in some key aspects. Because their origin is different they will depend differently on disorder and material properties such as SOC. Perhaps more importantly, the symmetry of the odd spin currents is distinct from the SHE. As a consequence the odd spin currents can have different spin-polarization than SHE, which could be important for the SOT \cite{MacNeill2016}. Furthermore, since these spin currents are odd under time-reversal, they will tend to cancel out in samples with many magnetic domains. Recently several experiments have demonstrated a SOT in Mn$_3$Ir/FM heterostructures \cite{Tshitoyan2015,Zhang2016,Oh2016,Wu2016}. While the origin of such a torque is not clear \cite{Oh2016} it is known that in heavy metal/FM heterostructures, the SHE plays an important role \cite{Fan2014,freimuth2014}. Since our calculations show that the odd transverse spin currents are in Mn$_3$Ir larger than the intrinsic SHE (the intrinsic SHE in Mn$_3$Ir is 215 $\hbar/e (\Omega \cdot \text{cm})^{-1}$ \cite{YangZhang2017}), we expect them to also contribute to the SOT. Taking the odd spin currents into account could help towards a better understanding of the unexplained features of the SOT \cite{Oh2016}.

In conclusion, we have shown that novel spin currents occur in non-collinear AFMs and that as a consequence electrical current in these materials is spin-polarized. The spin-polarized current is analogous to the spin-polarized current in FMs and could be therefore utilized in the same way. This could have important implications for the field of AFM spintronics since several key spintronics phenomena are based on the existence of spin-polarized current. We show that\textemdash just like the AHE\textemdash the novel spin currents are a consequence of a symmetry breaking caused by the non-collinear magnetic structure. The conclusions we have made are quite general: the odd spin currents will be present in most magnetic materials except simple collinear AFMs.

\begin{acknowledgments}
We thank Yuriy Mokrousov, Jairo Sinova, Tom\'a\v{s} Jungwirth, Jacob Gayles and Karel V\'{y}born\'{y} for fruitful discussions. We acknowledge support from EU FET Open RIA grant no. 766566. (ASPIN). C.F. acknowledges the funding
support by ERC (Advanced Grant No. 291472 “Idea Heusler”).
Y.Z. and B.Y. acknowledge the German Research Foundation
(DFG) SFB 1143. 
\end{acknowledgments}

%


\end{document}